# A massive, dead disk galaxy in the early Universe

Sune Toft[1], Johannes Zabl[1,2], Johan Richard[3], Anna Gallazzi[4], Stefano Zibetti[4], Moire Prescott[5], Claudio Grillo[6,1], Allison W.S. Man[7], Nicholas Y. Lee[1], Carlos Gómez-Guijarro[1], Mikkel Stockmann[1], Georgios Magdis[1,8], Charles L. Steinhardt[1]

**At redshift z = 2, when the Universe was just three billion years old, half of the most massive galaxies were extremely compact and had already exhausted their fuel for star formation[1–4]. It is believed that they were formed in intense nuclear starbursts and that they ultimately grew into the most massive local elliptical galaxies seen today, through mergers with minor companions[5,6], but validating this picture requires higher-resolution observations of their centres than is currently possible. Magnification from gravitational lensing offers an opportunity to resolve the inner regions of galaxies[7]. Here we report an analysis of the stellar populations and kinematics of a lensed $z$ = 2.1478 compact galaxy, which—surprisingly—turns out to be a fast-spinning, rotationally supported disk galaxy. Its stars must have formed in a disk, rather than in a merger-driven nuclear starburst[8]. The galaxy was probably fed by streams of cold gas, which were able to penetrate the hot halo gas until they were cut off by shock heating from the dark matter halo[9]. This result confirms previous indirect indications[10–13] that the first galaxies to cease star formation must have gone through major changes not just in their structure, but also in their kinematics, to evolve into present-day elliptical galaxies.**

We obtained deep spectroscopy using the XSHOOTER instrument on the Very Large Telescope (VLT) of a compact quiescent galaxy that is gravitationally lensed by the z = 0.588 cluster of galaxies MACS J2129.4−0741 (hereafter MACS2129−1; ref. 14; (Extended Data Fig. 1), and as a consequence appears 4.6 ± 0.2 times brighter and extends over 3″ on the sky.
In Fig.1, we show the position of the XSHOOTER slit on a Hubble Space Telescope (HST) colour-composite image and on the reconstructed source plane. The galaxy is stretched along its major axis and we derive a spatially resolved spectrum typical of quiescent z ≈ 2 post-starburst galaxies, with a strong Balmer break and a number of strong absorption features. We fit a spectroscopic redshift of z = 2.1478 ± 0.0006 and constrain the stellar populations through modeling of the rest-frame ultraviolet-to-optical spectrum, the absorption line indices, and the spatially integrated rest-frame ultraviolet-to-near-infrared (NIR) colours derived from 16-band HST/Infrared Array Camera (IRAC) photometry. The best-fitting spectrum reveals a massive, old, post-starburst

[1]Dark Cosmology Centre, Niels Bohr Institute, University of Copenhagen, Juliane Maries Vej 32, København Ø, 2100, Denmark.
[2]IRAP, Institut de Recherche en Astrophysique et Planétologie, CNRS, 9, avenue du Colonel Roche, F-31400 Toulouse, France/Université de Toulouse, UPS-OMP, Toulouse, France, [3]Centre de Recherche Astrophysique de Lyon, Observatoire de Lyon, Université Lyon 1, 9 Avenue Charles André, 69561 Saint Genis Laval Cedex, France. [4]Istituto Nazionale di Astrofisica–Osservatorio Astrofisico di Arcetri, Largo Enrico Fermi 5, 50125 Firenze, Italy. Department of Astronomy, [5]New Mexico State University, 1320 Frenger Mall, Las Cruces, NM 88003-8001, USA, [6]Dipartimento di Fisica, Universitá degli Studi di Milano, via Celoria 16, I-20133 Milano, Italy, [7]European Southern Observatory, Karl-Schwarzschild-Straße 2, 85748 Garching bei München, Germany, [8]Institute for Astronomy, Astrophysics, Space Applications and Remote Sensing, National Observatory of Athens, GR-15236 Athens, Greece

galaxy consistent with negligible ongoing star formation, and at most solar metallicity (that is, $\log[M_*/M_\odot] = 11.15^{+0.23}_{-0.23}$, $\log[\mathrm{Age(yr)}] = 8.97^{+0.26}_{-0.25}$, $A_i = 0.6^{+0.9}_{-0.6}$ mag., $\log(Z/Z_\odot) = -0.56^{+0.55}_{-0.55}$) (Extended Data Table 1). We derive a velocity dispersion of σ=329±73 km s$^{-1}$ from absorption lines in the spatially integrated spectrum (Fig. 1) using a penalized pixel fitting method (pPXF)[15] with the best-fitting spectral energy distribution model as a template. Interestingly, the absorption lines are tilted in the two-dimensional spectrum. We extract individual rows that represent approximately 0.4-kpc bins along the major axis on the source plane. The central 11 rows have sufficient signal to noise (S/N) to detect absorption lines reliably. We fit each row with the same pPXF implementation used for the spatially integrated spectrum to derive velocity shifts and dispersions as a function of distance from the centre of the galaxy. There is a clear symmetric velocity gradient, as expected for a rotating disk (Fig. 2).

The observed rotational velocity reaches a maximum of |V$_{\mathrm{max,obs}}$(1'')|=341±115 km s$^{-1}$ (weighted mean in the two outer bins), showing unambiguously, and independently of model assumptions, that the galaxy has a higher degree of rotational support V$_{\mathrm{max,obs}}$/σ > 1.03 ± 0.42 than observed previously in any galaxy that has ceased star formation (quenched). This is a conservative lower bound, as corrections for the effects of inclination, alignment of the slit, and unresolved rotation will all drive the ratio up. We take these into account, as well as the effect of gravitational lensing, by performing a full Markov chain Monte Carlo dynamical modelling analysis. We find the velocity shifts to be well represented by the rotation curve of a thin, circular, rotating disk with a maximum rotational velocity of $V_{\max} = 532^{+67}_{-49}$ km s$^{-1}$ at $R_{\max} = 0.5^{+0.8}_{-0.3}$ kpc. This implies a dynamical mass within $r_e$ of $\log(M_{\mathrm{dyn}}/M_\odot) = 11.0^{+0.1}_{-0.1}$ (and a total dynamical mass of $11.3^{+0.1}_{-0.1}$).

We model the observed dispersions by taking into account the effect of point spread function (PSF) smearing $\sigma^2_{\mathrm{obs}}(r) = \sigma^2_{\mathrm{int}} + \sigma^2_{\mathrm{sm}}(r)$. This is particularly relevant in the central regions where the steep gradient of rotational velocity translates into artificially high apparent velocity dispersions. We find that σ$_{\mathrm{obs}}$(r) is well represented by velocity smearing (σ$_{\mathrm{sm}}$) alone, but can also accommodate a small intrinsic constant dispersion ($\sigma_{\mathrm{int}} = 59^{+57}_{-44}$ km $s^{-1}$). In the central bin, σ$_{\mathrm{obs}}$ is about 0.5 standard deviation above the modelling predictions, leaving room for enhanced central dispersion; however, as there is no indication of a bulge in the stellar mass map (Fig. 3), any bulge is unlikely to be very prominent. From the dynamical modelling we conclude that that the galaxy is a rotation-dominated disk with V$_{\max}$/σ$_{\mathrm{int}}$ > 3.3 (97.5% confidence).

We detect weak, centrally concentrated nebular emission lines [O III] at wavelength λ = 5,007 Å, He II (5,411 Å), Hα (6,563 Å), and [N II] (6,583 Å). The (extinction-corrected) Hα flux corresponds[16] to a star-formation rate of SFR < 1.1 M$_\odot$ yr$^{-1}$, an upper bound since the line ratios indicate an active galactic nucleus (AGN) or a low-ionization nuclear emission-line region (LINER) as the dominant ionizing source[17], consistent with what is found in local post-starburst galaxies[18]. The emission lines are redshifted by a systematic velocity shift of V$_{\mathrm{el}}$ = 236 km s$^{-1}$ with respect to the absorption lines, possibly owing to AGN outflows (see Extended Data Fig. 2 and Extended Data Table 1).

Using our well constrained lens model and multi-band HST imaging, we reconstruct the galaxy on the source plane and derive spatially resolved maps of median-likelihood estimates of stellar population parameters (Fig. 3). The projected mass distribution is smooth, ruling out the possibility of a close major merger causing the velocity gradient.

The three other maps show indications of radial gradients, with the galaxy centre having a specific SFR about 2 dex lower, stellar populations about 0.2 dex older, and extinction about 0.6 mag higher than the outer regions (see Extended Data Fig. 3). Note, however, that the specific SFR is consistent with zero everywhere (see error map).

The quiescent nature of MACS2129−1 is further supported from its non-detection in deep Spitzer/Multiband Imaging Photometer for Spitzer (MIPS) 24-µm observations, corresponding to a lensing-corrected 3σ upper limit of SFR < 5$M_\odot$ yr$^{-1}$.

We fit the two-dimensional surface brightness distribution of the galaxy on the source plane. Figure 4 shows the reconstructed F160W image, the best-fitting Sersic model (circularized effective radius $r_e=1.73^{+0.34}_{-0.27}$ kpc, Sersic index n =$1.01^{+0.12}_{-0.06}$, axis ratio a/b = $0.59^{+0.03}_{-0.09}$) and the residual (image minus model), which shows visual hints of spiral arm structure. Also shown are one-dimensional profiles of the best-fitting model (black curve), the F160W light distribution (red curve) and the projected stellar mass distribution (grey area) estimated from the spatially resolved spectral energy distribution fits. The galaxy is resolved on scales of about 250 pc, allowing accurate constraints on its inner profile to around $r_e$/5. The light and mass profiles are well represented by an exponential disk law (n = 1) out to 5 kpc (approximately 3$r_e$), as expected if the galaxy is a rotation-dominated disk. The outer regions (r > 3re) have excess low surface brightness light compared to the model, which is not associated with notable excess stellar mass.

MACS2129−1 presents a range of features typical of z > 2 compact quiescent galaxies: it falls on the stellar mass–size relation for quiescent galaxies[2], it has a post-starburst spectrum with evolved stellar populations, a SFR more than 100 times below the z = 2 SFR–M∗ relation[19], and $M_{dyn}/M_* = 1.55 \pm 0.40$ (within $r_e$), similar to what is found in other z = 2 quiescent galaxies[3,4,13]. However, the increased sensitivity and resolution provided by gravitational lensing reveal underlying detailed kinematic properties and structure that are typical of late-type galaxies. The V/σ is > 3 times larger than in the fastest-rotating local ellipticals[20], but is similar to what is found in local spiral galaxies[21] (see Extended Data Fig. 4).

The only other kinematic study of the inner regions of the lensed z > 2 compact quiescent galaxy (RG1M0150)[7] revealed a dispersion-dominated (V/σ=0.7±0.2) proto-bulge (n=3.5±0.9), as expected if it formed in a major merger-induced nuclear starburst[8].

In sharp contrast, MACS2129−1 shows no evidence of a bulge, and must have formed through other processes. Rather than in a merger, its stars formed rapidly in a disk until quenching at z ≈ 3, which, based on the specific SFR–age profiles, proceeded inside-out over a timescale of about 300 million years. Simulations show that collisions of cold gas streams with the inner disk (and with each other), can trigger isothermal shocks, behind which rapid cooling generates intense star formation[7]. This can lead to the build-up of a central bulge that stabilizes the disk against further fragmentation, "morphologically quenching" and transforming the galaxy from the inside out[22]. If the star formation in MACS2129−1 was fuelled by cold accretion, the gas must have distributed itself throughout the disk, rather than in the centre, so a different mechanism would be required to quench its star formation. Shock heating of the in-falling gas by the halo is a plausible scenario, given the total halo mass of $10^{13}M_\odot$ – $10^{14}M_\odot$ indicated by its measured stellar mass[23]. 'Halo quenching' is expected to cut off the raw material for star formation from the inside out (as the critical halo mass for shock stability in the centre is smaller than at the virial radius[9]), but otherwise leaves the structure of the galaxy untouched. This is in agreement with the dynamics and stellar population gradients observed in MACS2129−1. The AGN can also play a part in maintaining the hot gas temperature, as AGN feedback works very efficiently on dilute, shock-heated gas[24].

Although the difference between MACS2129−1 and RG1M0150 is striking, both have characteristics consistent with their being descendants of the highest-redshift starbursts[5], which show diverse properties. Their molecular gas reservoirs are in some cases distributed in compact fast-spinning disks[25], while in others they are centrally concentrated and dispersion dominated[26]. Immediately after quenching, the starbursts probably go through a transition phase where they appear as $2 < z < 3$ dusty compact star-forming galaxies[27]—some of which also show evidence of rapid rotation[28]—before evolving into compact quiescent galaxies.

To evolve into a local elliptical, MACS2129−1 must undergo major structural and kinematic changes, through processes capable of both substantially increasing its size[2] and transforming the spatial and orbital distribution of its stars from ordered rotation in a disk to random motion in a bulge. The same is to some extent true for RG1M0150.

Although it is already a dispersion-dominated bulge galaxy, its $V/\sigma$ is much higher than similar-mass local elliptical galaxies, which are all slow rotators ($V/\sigma < 0.2$). Simulations show that a large number of minor mergers from random directions are probably the dominant process[29].

**Online Content** Methods, along with any additional Extended Data display items and Source Data, are available in the online version of the paper; references unique to these sections appear only in the online paper.

**Acknowledgements** S.T., J.Z., G.M., N.Y.L., C.L.S., C.G.-G. and M.S. acknowledge support from the ERC Consolidator Grant funding scheme (project ConTExt, grant number 648179). C.G. acknowledges support from the VILLUM FONDEN Young Investigator Programme (grant number 10123). G.M. acknowledges support from the Carlsberg Foundation and from the VILLUM FONDEN Young Investigator Programme (grant number 13160). S.Z. and A.G. acknowledge support by the EU Marie Curie Career Integration Grant "SteMaGE" number PCIG12-GA-2012-326466 (call identifier FP7-PEOPLE-2012 CIG). J.Z. acknowledges support of the OCEVU Labex (ANR-11-LABX-0060) and the A* MIDEX project (ANR-11-IDEX-0001-02) funded by the 'Investissements d'Avenir' French government programme managed by the French National Research Agency (ANR). We thank M. Yun and R. Cybalski for providing the deep Spitzer data, and D. Watson and F. Valentino for discussions.

**Author Contributions** S.T. conceived the study, was the Principal Investigator of the XSHOOTER programme, performed the Galfit analysis and produced Figs 2–4 and Extended Data Figs 3, 4 and 6. S.T. and J.Z. wrote the paper. J.Z. reduced the XSHOOTER data, performed the pPXF analysis and lensing model systematic error analysis. J.Z. also produced Fig. 1 and Extended Data Figs 5 and 7. A.G. performed the stellar population synthesis modelling of the spectrum and photometry. S.Z. performed the emission line analysis, produced the resolved stellar population maps and Extended Data Fig. 2. J.R. performed the lensing analysis, and source plane reconstruction. M.P. performed the Markov chain Monte Carlo dynamical modelling and produced Extended Data Fig. 8. C.G. produced the colour composite HST images in Fig. 1 and Extended Data Fig. 1. A.W.S.M. performed the Galfit Markov chain Monte Carlo analysis. G.M. derived the SFR limit from the MIPS data. All authors discussed the results and commented on the manuscript.

**Author Information** Reprints and permissions information is available at www.nature.com/reprints. The authors declare no competing financial interests. Readers are welcome to comment on the online version of the paper. Publisher's note: Springer Nature remains neutral with regard to jurisdictional claims in published maps and institutional affiliations. Correspondence and requests for materials should be addressed to S.T. (sune@dark-cosmology.dk).

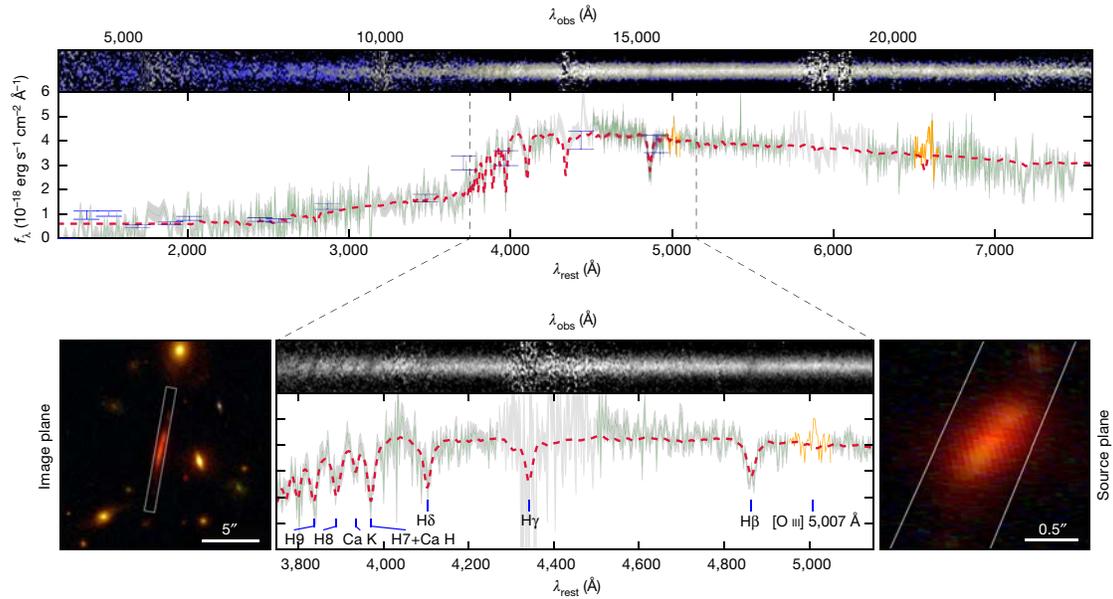

**Figure 1 | Spectrum of MACS2129−1.** Rest-frame ultraviolet–optical two-dimensional and one-dimensional XSHOOTER spectra, adaptively rebinned to a constant S/N per bin. The flux density ($f_\lambda$) is plotted as a function of the observed wavelength ($\lambda_{obs}$) and the rest-frame wavelength ($\lambda_{rest} = \lambda_{obs}/(1 + z)$, where z is the redshift). The bottom panel shows a close up of the most important absorption features, binned to a resolution of 9.6 Å (observed). The dashed red line shows the best-fitting stellar population model. Grey regions indicate windows of high telluric absorption; orange regions indicate important emission lines. The insets show colour-composite HST images of the galaxy on the image plane and on the reconstructed source plane, with the position of the slit overlaid.

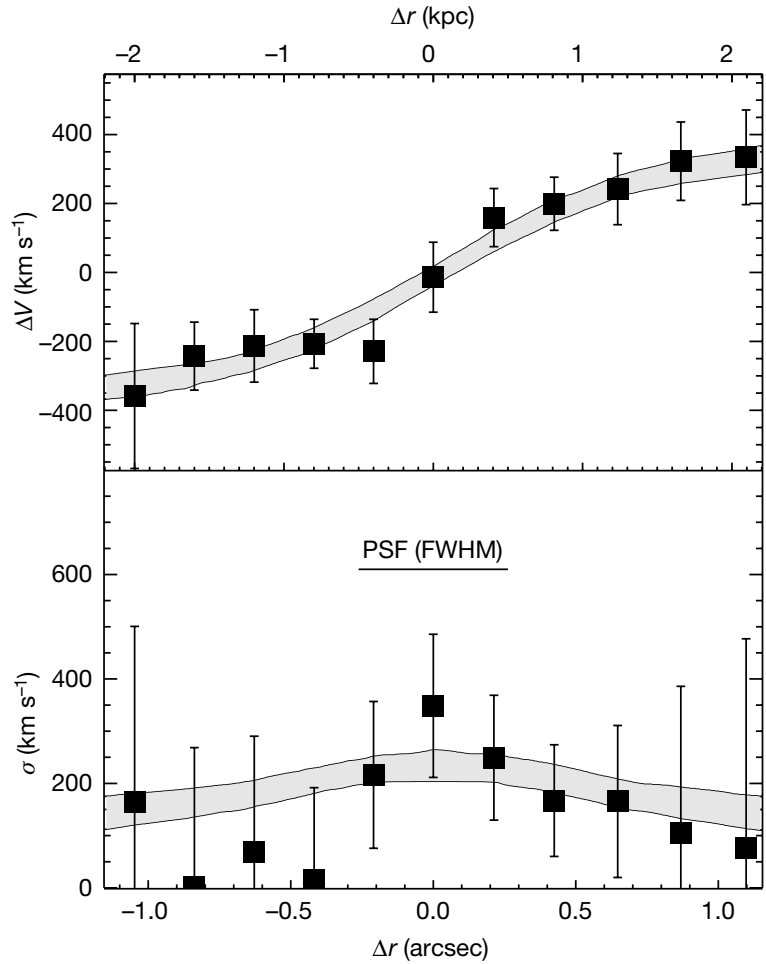

**Figure 2 | Rotation and dispersion curve for MACS2129−1**. Velocity offsets Δ V and dispersions σ as a function of distance from the centre of the galaxy Δ r, derived from pPXF fits to the individual spatial lines in the full spectrum. Error bars represent statistical uncertainties associated with the fitting and systematic uncertainties added in quadrature. The grey shading shows the 68% confidence interval for the thin disk model fits to the black squares. The observed dispersion is high in the centre and drops off symmetrically with distance, consistent with the effect of point spread function (PSF) smearing of the velocity gradient, and a constant dispersion of up to about 100 km s$^{-1}$ (grey shading). PSF (FWHM) indicates the full-width at half-maximum of the PSF.

$11.15^{+0.23}_{-0.23}$    $8.97^{+0.26}_{-0.25}$    $0.6^{+0.9}_{-0.6}$

$0.56^{+0.55}_{-0.55}$

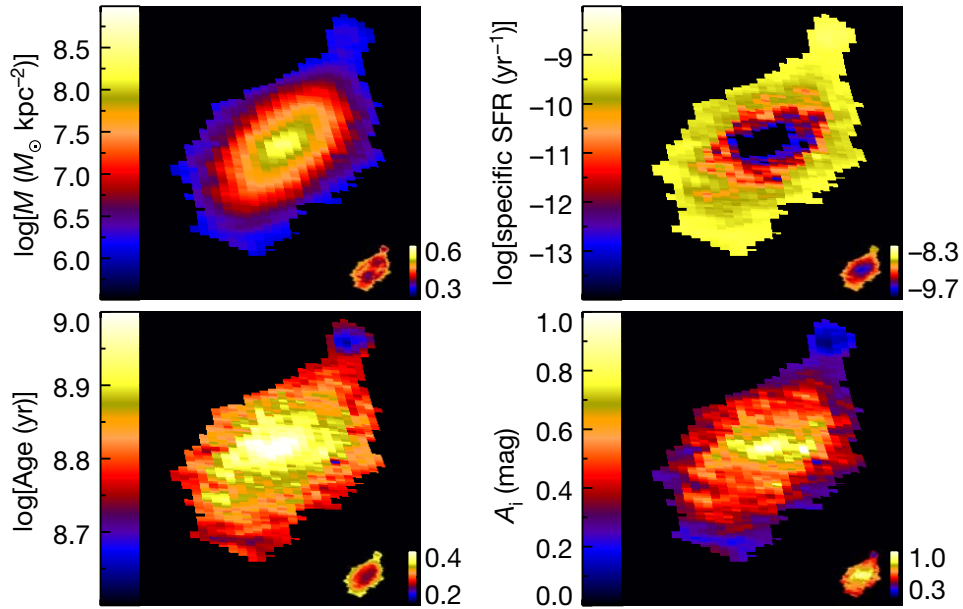

**Figure 3 | Stellar population maps on the reconstructed source plane.** Shown are two-dimensional maps of the stellar mass surface density M, the specific SFR, the stellar age and the dust extinction in the rest-frame i-band, $A_i$. These are created from fits to the HST imaging, using the same stellar population library as used to fit the full spectrum. The insets at bottom right show 68% confidence intervals for the derived parameters. The PSF is shown in Fig. 4. The younger knot in the top right corner, which may be an ongoing minor merger, does not give rise to the extra light seen in Fig. 4, or influence the conclusions based on azimuthally averaged profiles (see Methods).

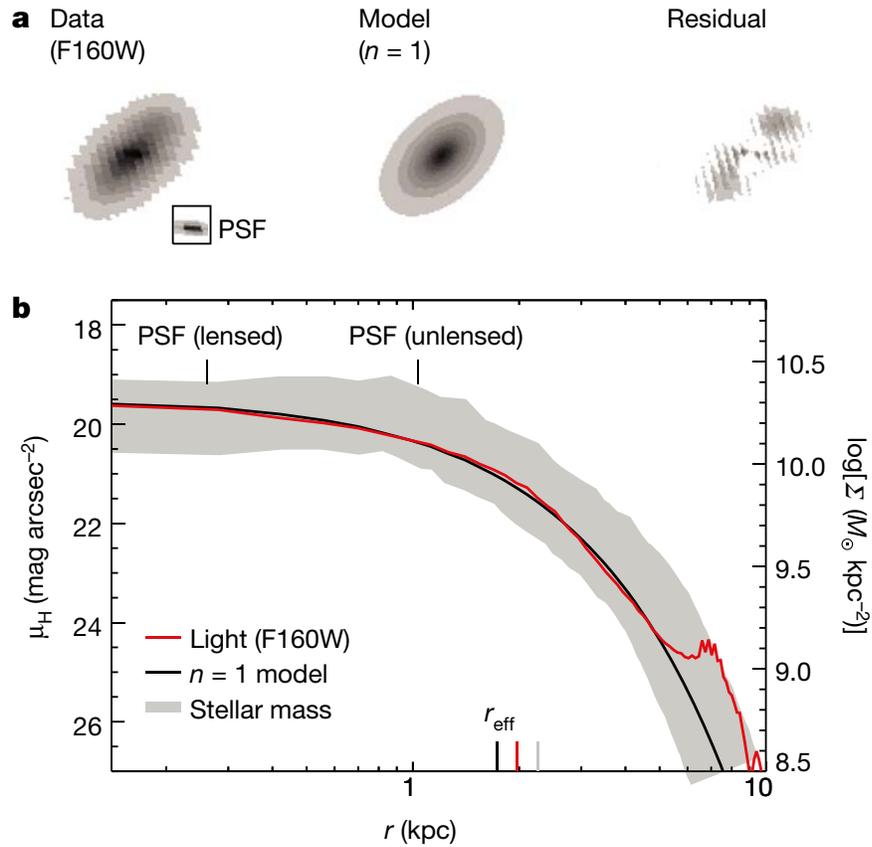

$1.01^{+0.12}_{-0.06}$

**Figure 4 | Surface brightness and stellar mass surface density profiles for MACS2129−1.** a, Left, the source-plane reconstruction of the WFC3/F160W band image with the PSF shown as an inset. Centre, the best-fitting Sersic model (n =$1.01^{+0.12}_{-0.06}$). Right, the residual, which may show a hint of spiral structure. b, The one-dimensional surface brightness ($\mu_H$) and stellar surface mass density ($\Sigma$) profiles (derived in elliptical apertures, following the best-fitting Sersic model) are both well represented by an n = 1 exponential disk model. The increased spatial resolution due to lensing is illustrated by the (circularized) half-width at half-maximum (HWHM) of the PSF at z = 2.15, with and without lensing.

$1.73^{+0.34}_{-0.27}$  $1.01^{+0.12}_{-0.06}$  $0.59^{+0.03}_{-0.09}$

**Methods**

**Selection of MACS2129−1.** MACS J2129.4−0741 at $z = 0.588$ was first listed in the Massive Cluster Survey (MACS) catalogue[14]. It was observed as part of the CLASH survey[30] with the HST/ACS and WFC3 in 16 broadband filters, from the near-ultraviolet to the near-infrared, and with Spitzer/IRAC in the 3.6-μm and 4.5-μm channels. Lensing models for the cluster have been developed and presented[31–33]. In Extended Data Fig. 1 we show a composite-colour image constructed from the CLASH data, with the position of the XSHOOTER slit overlaid on MACS2129−1.

**VLT/XSHOOTER observation and reduction.** MACS2129−1 was observed with the XSHOOTER spectrograph at the European Southern Observatory (ESO)'s VLT in four different observing blocks, executed on the nights starting on 11 March, 13 March, 4 September and 13 September 2011, respectively (programme ID: 087.B-0812(B)). Each observing block contained six integrations of 480s each, identical in all three of XSHOOTER's arms (UVB, VIS, NIR), and the slit widths were 1.0″, 0.9″ and 0.9″ respectively. The slit was oriented along the maximum extent of the image of MACS2129−1 with a position angle of −10°, as shown in Fig. 1. Photometric standard star observations of LTT7987, LTT3218, Feige110 and LTT7987 were taken during the same nights as the science observations. For each of the observing blocks, at least one telluric star was observed at similar air mass and with the same instrumental setup as the science frames. The average seeing was about 0.5″ in the J-band, estimated based on header information provided by the observatory which we calibrated based on point sources (telluric stars) to direct FWHM measurements. This seeing estimate is somewhat uncertain and we estimate an uncertainty of about 0.1″.

An analysis of this dataset has previously been published[34]. Here we take advantage of notable advances in data reduction and analysis techniques to revisit the observations. The ESO XSHOOTER pipeline (v. 2.6.0) with small modifications[35], was used in the reduction. In the near-infrared we achieved the best reduction with a nodding reduction, which we, for consistency, used also for the UVB and VIS data.

The offsets between the individual exposures were not ideally chosen for a nodding reduction of an object as extended as MACS2129−1. Whereas for half of the observing blocks a nod throw of 3″ was used, which is not ideal but acceptable, for two of the four observing blocks consecutive frames had very small position differences, rendering these pairs useless for a standard nodding reduction. As a solution, we combined frames in these observing blocks in a not strictly temporal order, and rejected two frames in each. In this way we obtained for each of these two observing blocks one pair with a 3″ and one with a 2″ nod throw. Excluding the rejected frames we had a total on-source time of

2.67 h.

We reduced the data in pairs of two frames and corrected each of the reduced nodding pairs for the telluric transmittance[36]. We then corrected from air to vacuum wavelength scale and converted to the heliocentric velocity standard. The latter step is relevant, as the heliocentric velocity differs by as much as 44 km s$^{-1}$ between the March and September frames. As the difference in the heliocentric velocity is comparable to the spectral resolution (about 55 km s$^{-1}$), this correction is a very efficient way to minimize the impact of telluric features. Finally, we combined all nodding pairs using a weighted mean. Further, we created re-binned versions of the two-dimensional spectrum and extracted one-dimensional spectra from the two-dimensional frames within the various apertures used in the analysis.

**Lensing model and source plane reconstruction.** The mass distribution of the cluster core was modelled using the following strong lensing constraints: a $z = 1.363$ source producing six images near the central galaxies[37] and a triply imaged system at $z = 2.26$ on the west side of the core[38]. We used these nine images to constrain a parametric model of the cluster, composed of a single large-scale pseudo-isothermal elliptical profile as well as sub-structure in the form of individual potentials for each galaxy in the cluster. Cluster members were selected through their HST colours and assumed to follow a scaling relation with constant mass-to-luminosity ratio $M/L$, except for two galaxies near the cluster cores which affect the location of the images in the $z = 1.363$ system.

The optimization of the cluster mass distribution was done with the Lenstool software[38], which uses a Monte Carlo Markov chain to provide a best-fitting model together with approximately 2,000 model realizations sampling the posterior probability distribution of each parameter of the mass distribution, including the uncertainty in the $M/L$ ratio of each of the cluster galaxies. We used the best-fitting model as a benchmark to perform a reconstruction of all the different wave-band HST images by mapping the image plane maps onto the source plane and associating each image pixel to its respective source-plane position.

To estimate the effects of lensing model uncertainties on the analysis, in Extended Data Fig. 5 we show distributions and correlations between parameters that characterize the lensing model realizations. These are the mean light (F160W)-weighted magnification ($\mu_{lw}$), the direction of maximum magnification at the centre of the galaxy, and the amount of magnification in this direction and the perpendicular direction. The variations of these quantities are small; for example, the mean light-weighted magnification displays only mild variations over the realizations $\mu_{lw} = 4.6 \pm 0.2$. Even the most extreme magnification values (4.2 and 5.6 respectively) deviate less than 20% from the median. Similar conclusions are drawn for the direction of the magnification, measured at the centre of the galaxy.

Lensing model uncertainties propagate into the measurement of structural and kinematical parameters, as also demonstrated in Extended Data Fig. 5. For example, the derived galaxy axis ratio and position angle, which are used as priors in the kinematical analysis, are strongly correlated with both strength and orientation of the lensing. However, as the lensing model uncertainties are small, uncertainties on structural parameters derived from Galfit are also small (see also Extended Data Fig. 6).

Finally, to demonstrate the validity of the use of a constant lensing kernel for simulating the seeing in the dynamical modelling, we quantified the spatial gradients of the scalar magnification over the galaxy in Extended Data Fig. 7, both for a typical realization, and for the realizations with the highest and lowest lensing magnifications. There is some variation caused by a cluster galaxy 3.5″ to the west. The impact at the position of the MACS2129−1 is, however, relatively small: over the full extent of the galaxy the magnification changes by about 30%, but within the spatial bins defined by the slit and the spatial binning used for the kinematic measurement the light-weighted magnification (taking into account the seeing), does not change by more than 5%.

The reason that the model magnification is so stable for MACS2129−1 is its position 1 arcmin west of the core, outside the strong lensing region where the magnification is high and sensitive to mass substructures associated with the high density of cluster galaxies. Outside the core, the lensing magnification is dominated by the diffuse dark-matter distribution of the cluster, which is well constrained by the multiply imaged galaxies. A good level of precision and accuracy in the reconstruction of images far from the critical curves is a general result, which has been established within the HST Frontier Fields initiative, through detailed comparison of blind predictions of several modellers on simulated galaxy clusters[39].

**Spectral stellar population characterization.** We constrained the global stellar mass, age, metallicity and extinction of MACS2129−1 by fitting the spatially integrated ultraviolet–NIR XSHOOTER spectrum with stellar population synthesis models, following the procedure described in ref. 4 but adopting a different model library that better reproduces the current star-formation activity.

In brief, we use a library of 500,000 model spectra based on BC03 stellar population models convolved with random star-formation histories. We assume a Chabrier initial mass function[40]. The star-formation histories are modelled with a delayed exponential[41], on top of which random bursts of star formation can occur (library A). We also consider a star formation history library with no additional bursts (library B). The metallicity of the models varies along with the star- formation history and dust is implemented as in ref. 42. We adopt a Bayesian approach in which all models in the library are compared to the observed spectrum. Details of the adopted star-formation history, metallicity and dust library as well as the method used have previously been published[43].

As a benchmark we fit the UVB–VIS–NIR spectrum pixel by pixel (fit 1), but as a test of the robustness of the fit we also perform a fit to observed spectral indices: D4000n, H$\beta$, H$\delta_A$, [Mg$_2$Fe], [MgFe]′, combined with optical–NIR (CLASH) and mid-infrared (MIR) (IRAC) broadband data (fit 2). The marginalized median- likelihood parameters are listed in Extended Data Table 1. Reassuringly, these all agree to within estimated uncertainties. The galaxy is massive, old, and within uncertainties consistent with being void of dust and star formation. The metallicity is quite uncertain, but has an upper confidence limit of approximately solar in all fits. The large uncertainties on the metallicity are reflected in the uncertainties of the other parameters compared to fits assuming a fixed metallicity[3], but we choose to keep it a free parameter in the fit, because we do not have much empirical evidence for the metallicity in $z = 2$ quenched galaxies. In addition to fitting the global properties in the spatially integrated spectrum, we also consider a fit to a 'central' ($|r|<0.5''$), and an 'outer' ($0.5'' < |r| < 1.4''$) extraction to look for spatial variations.

**Derivation of rotation curve.** We used pPXF (v. 6.0) to determine the velocity centroid (redshift) and dispersion both for the integrated spectrum and individual spatial rows of the rectified two-dimensional spectrum. The individual rows correspond to a spatial extent of 0.21″ each. Because the individual rows have low $S/N$ for kinematic fitting (central row $S/N = 3$ Å$^{-1}$ at 1.05″ $S/N = 1$ Å$^{-1}$), it is crucial to minimize the degrees of freedom and maximize the amount of information. In this sense, we derive our fiducial results based on the best-fitting stellar population fit and used only low-order correction polynomials (second-order additive and multiplicative over the full rest-frame range from 3,750 Å to 5,650 Å, excluding only a window from 4,200 Å to 4,750 Å (the gap between the J and H bands) and the [O III] ($\lambda = 4,959$ Å and 5,007 Å) lines. Certainly, the inclusion of the 4,000 Å break combined with low-order correction polynomials is a potential concern for template mismatch. We quantified potential template mismatch as in other studies[3,13] through a range of tests including different correction polynomials, different templates, and different wavelength ranges. The full error budget includes both statistical and systematic uncertainties.

We determined the statistical uncertainties by running pPXF on 1,000 realizations of the data. These realizations were created by randomly perturbing the best-fitting model obtained by pPXF using the estimated uncertainties from the pipeline's error spectrum. This method correctly accounts for the increased noise in regions of telluric emission and absorption lines, if the error spectrum is correct. In this sense, correlated noise is a potential problem, as it is not accounted for by the pipeline. However, the correlation is mostly removed in the re-binned spectra used for our analysis. However, estimated per pixel uncertainties obtained through standard error propagation from the non-binned error spectrum is a wavelength-dependent under-estimation owing to correlation at the original pixel scale. We therefore decided to scale a propagated error map to a reduced $\chi^2$ of one. To do so, we first calculated a running reduced $\chi^2$ over 50 pixels and fitted a second-order polynomial to the resulting wavelength-dependent curve, making use of clipping against outliers. By dividing the formal error spectrum by this fit, we obtained the corrected error spectrum. To assess systematic uncertainties that are primarily due to template mismatch, we performed several tests on the integrated spectrum. For this spectrum the $S/N$ is sufficiently high ($S/N = 6.4$ Å$^{-1}$) that the tests are feasible without being completely dominated by statistical uncertainties.

First, we assessed the impact of the choice of correction polynomials when using our fiducial best-fitting stellar population template. We tested all combinations of multiplicative ($m$) and additive ($a$) correction polynomials up to 24th order each, finding a minimum and maximum dispersion of 271 km s$^{-1}$ and 361 km s$^{-1}$, obtained for $m = 0/a = 24$ and $m = 7/a = 7$ polynomials, respectively. The median dispersion from all polynomial combinations is very close (334 km s$^{-1}$) to the value obtained for our default correction polynomial set ($m = 2/a = 2$). It is reassuring that our chosen set of low-order polynomials gives a typical result.

As a second test, we checked the impact of changing the wavelength range included in the fit, while keeping the low-order correction polynomials and the fiducial template. We increased the start wavelength range from 3,750 Å up to 4,050 Å in steps of 5 Å. At 4,050 Å only Hδ and Hβ are left in the wavelength interval outside the telluric absorption windows. Similarly, we incrementally decreased the end wavelength from 5,650 Å down to 4,050 Å, again in steps of 5 Å. At 4,050 Å Hδ, Hβ are excluded and only the higher Balmer and the Ca H+K lines remain. Finally, we tested excluding windows 1,600 km s$^{-1}$ wide with the range of tests covering the complete included wavelength range (in 5-Å steps). There is some indication that the more the information is dominated by Hβ, the higher is the preferred dispersion. First, the exclusion test gives the lowest dispersion when Hβ is excluded (280 km s$^{-1}$). Second, the dispersion rises slightly with increasing the start wavelength.

The lowest dispersion is obtained when setting the start wavelength to 3,775 Å (320 km s$^{-1}$) and the highest dispersion value is reached for a start wavelength of 4,030 Å (400 km s$^{-1}$). When changing the end wavelength, the lowest velocity dispersion is reached for an end wavelength of 4,085 Å (290 km s$^{-1}$). Such a change of the measured velocity dispersion as a function of the included wavelength interval can also be a consequence of including or excluding the 4,000 Å break. In a previous study, where a similar trend was observed for other galaxies, this has been attributed to template mismatch in the 4,000 Å break[3].

Finally, we tested the impact of the choice of template with two different template sets. In one case we allowed pPXF to choose freely from BC03 simple stellar populations with solar metallicity, and in a second case from 45 stars that are a subset of the Indo-US library[44]. The stars were chosen[45] to cover the temperature space approximately uniformly between $3 \times 10^3$ K and $42 \times 10^3$ K and to have metallicities in the range $-0.5 <$ [Fe/H] $< 0.4$. For the tests with these template sets we used correction polynomials of higher order ($m = 2/a = 9$; about one order per 1,000 km s$^{-1}$) than used for the best-fitting stellar population template. Higher-order polynomials are needed for these template sets, for example, to account for dust extinction. We measured velocity dispersions of 403 km s$^{-1}$ and 386 km s$^{-1}$ based on the BC03 and Indo-US libraries, respectively.

To sum up the results of these systematic tests, we find variations, which are of the same order as the statistical uncertainty, which is 58 km s$^{-1}$ for the integrated spectrum. It is important to note that we had to use the data itself for the systematic tests. Therefore, the observed wavelength trend can be due to template mismatch, to lines in the 4,000 Å break or to the Hβ line, or simply to statistical noise altering the shape of the individual lines differently. In this sense, the systematic uncertainties are a conservative estimate, as they are not fully independent from the statistical uncertainties.

We quantified the full systematic error budget as $0.68(\sigma_{max} - \sigma_{min})/2$, where $\sigma_{max}$ and $\sigma_{min}$ are the maximum and minimum $\sigma$ obtained from all tests above (271 km s$^{-1}$ and 403 km s$^{-1}$). The total uncertainties used for the analysis were then derived by adding statistical and systematic uncertainties in quadrature. The uncertainties for the velocities (redshifts) were estimated in the same way.

We used the systematic uncertainties derived for the integrated spectrum also for the individual spatial extractions, because we expect the template mismatch uncertainties to be of similar extent. In addition, to make sure that the result does not depend on a single line (for example, owing to residuals), we repeated the test excluding individual wavelength intervals 1,600 km s$^{-1}$ wide in the fit of the individual spatial extractions. The uncertainties from this test were added additionally in squares.

**The tilted absorption lines are not artefacts of the reduction.** Owing to the relatively small nod throws, the object itself can potentially contaminate the region used for background subtraction, especially in the outer parts of the galaxy. Therefore, we simulated the impact of the nodding strategy on the spatial profile and on the rotation curve. From this we conclude that we can safely use the rotation curve out to about 1″.

As XSHOOTER is an echelle spectrograph, the geometrical mapping between the non-rectified and the rectified frame could potentially cause systematic effects, but we do not find any evidence of this. The residuals between predicted and actual positions of calibration lamp images do not deviate by more than one pixel and skylines are straight in the rectified output frames.

Another possible complication is the non-constant dark level of the NIR detector, especially variations between different rows of the detector, which trans- forms into half-circle-like noise structures in the rectified frames. At the low $S/N$ levels of the absorption lines, this may potentially conspire to make straight lines appear tilted. We simulated the appearance of straight lines on the detector in the rectified frames and did not find a clear correlation between the position of the half-circles and the observed absorption lines, arguing against this effect causing the tilt. Finally, the ($\tanh(x)$) rectification kernel could potentially cause problems; however, testing for this effect on several observed objects, we found important effects only for point sources observed under very good seeing conditions.

**Dynamical modelling.** We model the rotation curve using a thin disk model with seven parameters[46]: the offset angle between the slit and the major axis of the disk ($\Theta_{off}$), the disk inclination ($i$), the maximum velocity of the disk ($V_{max}$), the radius at which the disk reaches $V_{max}$ ($R_{max}$), the position of the centre of the slit relative to the disk centre ($X_c$, $Y_c$), where a positive offset in both parameters indicates a slit centre located to the northwest of the disk centre, and an intrinsic velocity dispersion ($\sigma_{int}$), assumed to be constant across the disk.

We follow a standard Markov chain Monte Carlo fitting approach with 100,000 iterations to find the combined best fit to the velocity and dispersion profiles. The fit uses a Metropolis–Hastings algorithm in which the proposed new value of each parameter is drawn from a Gaussian distribution centred at the current value and with a width drawn randomly from a specified range for each parameter. The allowed range was set empirically for each parameter to sample the posterior well while converging efficiently to a solution. The acceptance ratio $\alpha$—defined as the ratio between the computed likelihoods for the proposed and previous models— was used to accept or reject the proposed model at each step. If $\alpha$ was greater than 1, indicating that the proposed model was a better fit than the previous model, the proposed model was accepted; otherwise, the proposed new model was accepted only if $\alpha$ was greater than a random number drawn from a uniform distribution between 0 and 1. All parameters were varied simultaneously during the fitting and the first 10% of the iterations were treated as a burn-in period, and excluded from the final results.

During the fitting routine, each source plane model was convolved with a kernel that approximates the effects of seeing and lensing, weighted by the observed light profile of the galaxy and binned in the same way as the data. The dispersion measured for each model therefore includes the effects of velocity smearing as well as the intrinsic velocity dispersion, added in quadrature.

The single observed slit position angle limits our ability to break degeneracies among the parameters. However, our Galfit modelling provides firm constraints on the position and orientation of the galaxy with respect to the slit on the source plane, so we implement these as priors in the fit ($\Theta_{off}$ =22°±10°, $|X_c|$)<0.4 kpc. To speed up the fitting process, we also restrict the inclination to be within the range [45°, 60°]. The posterior distributions for each parameter are shown in Extended Data Fig. 8 as open histograms, along with the implied $V_{max}/\sigma_{int}$ and dynamical mass ($M \approx V^2 r /G$), where $r$ is the measured effective radius. The filled histograms show the subset of fits with inclinations within 3$\sigma$ of the inclination implied by the measured axis ratio, assuming the disk is circular when viewed face-on. We adopt these distributions as our best fits, and list the median values and 68% confidence intervals for each parameter of the sampled models in Extended Data Table 2.

To test the effect of the assumed lensing and seeing kernel on the results, we repeated the analysis using four different kernels. Two represent the extreme ends of the magnification found among the 1,979 realizations, and two represent the best and worst seeing plausible for our data. The results, summarized in Extended Data Table 2, are consistent with our benchmark model and show that lens model and seeing uncertainties are sub-dominant for our analysis when compared with observational uncertainties.

**Spatially resolved stellar population characterization.** The exquisite multi-band HST coverage, combined with the increased depth and resolution provided by the lensing, offers an opportunity to study trends in the spatially resolved stellar populations. To ensure sufficient S/N for robust stellar population characterization, we include only the F850LP, F105W, F110W, F125W, F140W and F160W band images in the fits. The bluer bands are too shallow and noisy over most of the galaxy's extent to provide useful information. Next, we PSF-matched the images, and adaptively smoothed them, using ADAPTSMOOTH[47,48]. Smoothing masks were constructed for the reddest images (F125W, F140W and F160W), by requiring a minimum S/N = 10 and a maximum smoothing radius of 3 pixels. These three masks were then combined and used to smooth all six images in the same way.

We then performed stellar population characterization for each pixel in the matched, smoothed images, and constructed two-dimensional maps of stellar mass, age, extinction ($A_i$) and specific SFR based on the median of their likelihood distributions (using the same technique as used in the benchmark fit of the spectrum). These maps were then transformed to the source plane using our well constrained best-fitting lensing model (Fig. 3). Error maps were constructed from the 68% confidence intervals of the fits to the individual pixels. In an absolute sense these parameters are subject to rather large uncertainties (see insets in Fig. 3) owing to age/dust/metallicity model degeneracies; however, because individual spatial bins are independent (on scales larger than 3 pixels), the relative spatial structure is more robust, and the integrated properties are consistent with those derived from the full spectral fit.

In Extended Data Fig. 3 we show radial profiles of the azimuthally averaged stellar population parameters (solid line). These show evidence for radial gradients with older stellar populations, higher extinction and lower specific SFR in the centre than in the outer parts. The shaded areas represent the pixel-to-pixel scatter in the median values in the elliptical apertures, not taking into account the uncertainties on the individual estimates. The amplitudes of these regions provide a rough representation of the random errors on the parameters, that is, owing to measurement errors alone. (We note that this scatter slightly underestimates the real random error due to pixel-to-pixel correlation, for example, because of the smoothing.) Also shown are the median-likelihood parameters values for the two spectral extractions, which are consistent with the photometric fits.

Despite the large systematic uncertainties, trends in physical parameters are evident and should be considered robust under the hypothesis that there is no radial trend in the possible systematic bias affecting our estimates. Given the homogeneity of the observables throughout the radial extent of the galaxy and also the relatively small range in physical parameters, we have no reason to expect any radially varying systematic bias.

The gradients are not driven by the age/dust degeneracy because that would result in opposite slopes for age and dust profiles. Also shown is the average specific SFR gradient derived from adaptive optics observations of similar-mass $z = 2$ star-forming galaxies[22], which has been interpreted as evidence of inside-out morphological quenching. The specific SFR gradient in MACS2129−1 has a similar shape, but is depressed by a factor of >100. The observed age gradient suggests that quenching in

MACS2129−1 proceeded from the inside out, over a timescale of about 300 million years.

Inspection of the maps shows an off-centre clump, with a younger stellar population and less extinction than the rest of the galaxy. This may indicate an ongoing minor merger, but we cannot confirm this because it is not detected in the spectrum. Masking out the clump does not change the derived radial stellar population gradients much.

**Surface brightness fits.** We used Galfit[49] to fit the source plane F160W (rest-frame optical) two-dimensional surface brightness profile with a Sersic model. As a PSF model we used a point source, reconstructed on the source plane in the same way as MACS2129−1 (see inset in Fig. 2). In addition to the formal fitting errors, uncertainties in the lensing model used for source plane reconstruction must be taken into account. We do this by re-running Galfit on 98 different reconstructed images and matching PSFs, selected to sample the full range of lensing model realizations.

The derived parameters are very stable (see Extended Data Fig. 6), with mean and standard deviations of $r_e$ = 20.9 ± 0.6 pixels, $a/b$ = 0.59 ± 0.01, $r_{e,cir}$ = 1.73 ± 0.05 kpc, $n$ = 1.01 ± 0.01, position angle PA = −45.17° ± 0.96°, demonstrating that lensing model uncertainties do not much affect the shape and orientation of the galaxy on the source plane. The main potential systematic uncertainty on our lens modelling is associated with the implementation of substructure. As a conservative upper limit on this uncertainty, we repeated the Galfit analysis on two extreme sets of model realization: one where the nearby cluster galaxy has zero mass and one where it has twice the mass. In the first case we obtain $r_e$ = 24.9 ± 0.6 pixels, $a/b$ = 0.49 ± 0.01, $r_{e,cir}$ = 1.89 ± 0.03 kpc, $n$ = 0.89 ± 0.01, PA = −35.49° ± 0.38°, and in the second $r_e$ = 18.2 ± 0.5 pixels, $a/b$ = 0.62 ± 0.01, $r_{e,cir}$ = 1.55 ± 0.04 kpc, $n$ = 1.07 ± 0.01, PA = −58.0° ± 1.8°.

From these tests we derive conservative estimates of the total errors (statistical and maximum systematic added in quadrature) for the Galfit parameters: $r_e = 20.9^{+4.0}_{-2.8}$ pixels, $a/b = 0.59^{+0.03}_{-0.09}$, $r_{e,cir} = 1.73^{+0.34}_{-0.27}$ kpc, $n = 1.01^{+0.12}_{-0.06}$, PA $= -45.0^{+9.8}_{-12.8}°$. We also repeated the dynamical modelling with these models, but found no notable changes in the results. This analysis is sensitive only to the brightest, central part of the galaxy which is not very strongly affected by the substructure, and its error budget is dominated by the limited $S/N$ of the spectrum, and PSF smearing. Finally, we note that masking out the north-east clump does not noticeably change the two-dimensional fit or the one-dimensional profile. It is thus not the main source of the 'excess light' seen at large radii.

To test what structural parameters we would have derived for MACS2129−1 in the absence of lensing we repeated our Galfit analysis on a 'de-lensed' version of the source-plane F160W image. This image was created by re-binning the F160W source plane image to the pixel scale of WFC3, convolving it with the WFC3 PSF, scaling its brightness by the inverse of the magnification factor and inserting it into an empty region of the original F160W image. The best-fitting parameters $r_{e,cir}$ = 1.92 ± 0.02 kpc and $n$ = 1.02 ± 0.03 are similar, demonstrating that MACS2129−1 would have been identified as a compact exponential disk, even in the absence of lensing.

**Emission lines may imply AGN outflow.** We decompose the stellar continuum and nebular emission lines in the visible–NIR spectrum using pPXF+Gandalf[50]. Detected lines in the central and outer extractions are listed in Extended Data Table 1. The emission lines are centrally concentrated. To estimate the hardness of the ionizing field, we calculate the line ratios $\log([\text{N II}]/\text{H}\alpha) = -0.06 \pm 0.10$ and $\log([\text{O III}]/\text{H}\beta) > 0.45 \pm 0.08$. The latter is a lower bound, since H$\beta$ is undetected, and estimated assuming case B recombination (H$\alpha$/H$\beta$ = 2.9) and no extinction. These ratios are inconsistent with O and B stars being the main ionizing source, but consistent with the expectations from AGN or a LINER[17].

The emission lines have dispersions similar to those measured from the absorption lines ($\sigma_{em}$ = 382 km s$^{-1}$), but are redshifted by 238 km s$^{-1}$. Velocity offsets of this order are expected from bipolar AGN outflows. Usually these are blueshifted, but in some local Seyfert galaxies redshifted lines are observed, when the angle of the bicone outflow is small with respect to the main plane of the host galaxy[51]. Alternatively, the redshift could be caused by spatially unresolved patchy star formation or dust in the inner part of the disk, but since the emission is centrally concentrated and the line ratios suggest AGN origin, this is a less likely scenario.

An exotic interpretation of the weak redshifted emission lines is that MACS2129−1 is a high-redshift analogue of the so-called 'offset AGN', observed in a large fraction of local early-type or post-merger galaxies in over-dense environments[52]. The offsets in these are interpreted as being caused by dual black holes only one of which is accreting; this is a sign of a major merger in the recent past[52]. Simulations show that if the merger were gas-rich (gas fraction >0.5) it would have commenced 1–2 billion years earlier[53], and that the gas in the remnant under such conditions can quickly rearrange itself in a rapidly spinning central disk[54]. This is a plausible alternative formation scenario for the massive stellar disk observed in MACS2129−1, consistent with its mean stellar age (around 750 million years).

**Rotation may be common in high-redshift quiescent galaxies.** The exponential profile and late type kinematics of MACS2129−1 appear in tension with the commonly accepted picture derived from lower-resolution data, that high-redshift quiescent galaxies are predominantly dispersion-dominated[6] protobulges with de-Vaucouleurs-like profiles[55], a picture that provides a straightforward link to their low-redshift descendants[5]. However, there is a growing body of indirect evidence that quiescent galaxies may grow more disk-like and rotation-dominated with redshift. A number of studies have found a fraction of quiescent galaxies to have exponential disk-like surface brightness profiles, both at high[11,13,56,57] and intermediate redshifts[58]. Indeed, the best candidate low-redshift direct descendant of a $z$ = 2 compact quiescent galaxy was found to be a rapidly rotating exponential disk[59]. Also, it has been argued that the average observed axis ratios of $z \approx 2$ compact quiescent galaxies are consistent with the majority of them being disks[12].

Recently it was argued that dynamical mass of $z \approx 2$ quiescent galaxies with exponential disk like profiles, calculated from unresolved spectroscopy was higher than expected from their stellar mass, which was interpreted as evidence for unresolved rotation[13]. The resolved kinematics for MACS2129−1 allows us to test this interpretation. If we calculate the dynamical mass in the same way as this study $\log[M_{dyn}(\sigma_e)] = \beta(n)\sigma_e^2 r_{e,maj}/G$, where $\beta(n) = 8.87 - 0.831n - 0.024n^2$, $n$ is the Sersic index, $\sigma_e$ is the spatially integrated velocity dispersion, and $r_{e,maj}$ is the major axis effective radius, we find $\log[M_{dyn}(\sigma_e)/M_\odot] = 11.65 \pm 0.14$, a factor of 2 higher than the total dynamical mass derived from resolved kinematics $\log[M_{dyn,tot}(V_{rot})/M_\odot]=11.30\pm0.14$ (assuming that $\log[M_{dyn,tot}(V_{rot})] = 2\log[M_{dyn,r<r_e}(V_{rot})]$). In the framework of the abovementioned study[13], the implied $\log(M_*/M_{dyn}) = -0.39 \pm 0.37$ suggests a contribution from unresolved rotation of up to $V/\sigma \approx 3$; however, given the error bars, the result is not statistically significant. We note that assuming $\beta = 5$ and $r_{e,c}$ instead of $r_{e,maj}$ (as often done in the literature), brings the unresolved $[M_{dyn}(\sigma_e)]=5\sigma_e r_{e,c}/G=11.33\pm0.09$ into perfect agreement with the value derived from the resolved kinematics.

Gravitational lensing currently offers the best way to quantify directly the ubiquity of rotation in high-redshift quenched galaxies until we have the Extremely Large Telescopes. Ground-based adaptive optics observations with 8–10 m telescopes can in principle resolve un-lensed examples, but will struggle with sensitivity when the signal is spread out. The James Webb Space Telescope will have the sensitivity required, but not quite the desired spatial resolution.

So far, three gravitationally lensed quiescent galaxies have been studied spectroscopically, MACS2129−1 ($z = 2.1$), RG1M0150 ($z = 2.6$) and COSMOS 0050+4901 ($z = 2.8$). The first two show clear evidence of rotation, the third did not have sufficient $S/N$ for us to be able to tell[60]. Gravitational lensing of quiescent galaxies is a rare phenomenon, but with systematic investigation of the large number of massive clusters available from existing and future surveys, it will be possible to build statistical samples of quiescent galaxies with resolved spectroscopy, one galaxy at a time.

**SFR limit from Spitzer/MIPS.** We derive an upper limit on the SFR in MACS2129−1 from its non-detection by Spitzer/MIPS at 24 μm (Principal Investigator M. Yun). From the root mean square of the 24-μm map we derive a $3\sigma$ upper limit of $f_{24}$ μm = 45 μJy. We convert this to an upper limit on the total infrared luminosity of $(2.3 \pm 2.0) \times 10^{11} L_\odot$, where $L_\odot$ is the solar luminosity, by assuming various local (M82, Arp220) and high-redshift main-sequence and starburst templates from ref. 62. We then correct for magnification and derive an upper limit for the SFR of $<5 M_\odot$ yr$^{-1}$, using the Kennicutt[16] relation (modified to a Chabrier IMF).

**Code availability.** The code used to perform the stellar population characterization from the XSHOOTER spectrum and the multi-band photometry is not publicly available, since full user documentation and interface are not yet developed. The base simple stellar population models adopted are publicly available at http://www.bruzual.org/bc03/. The adopted star-formation history library and the Bayesian fitting approach are fully described in ref. 43. The code ADAPTSMOOTH[47,48], used to PSF-match and adaptively smooth the images, is available at http://www.arcetri.astro.it/~zibetti/Software/ADAPTSMOOTH.html. The codes, pPXF[15] in combination with GANDALF[50], used to decompose the stellar continuum and the nebular emission lines in the XSHOOTER spectrum are publicly available at http://www-astro.physics.ox.ac.uk/~mxc/software/. We also made use of pPXF with a simple wrapping script (available upon request), for measuring the kinematics of the galaxies. Cluster mass modelling and source reconstructions were made using the latest version of the software Lenstool, which is publicly available at the following web page: https://projets.lam.fr/projects/lenstool/wiki. The parameter file for Lenstool is available upon request. The code used to perform the Markov chain Monte Carlo disk fitting analysis is not publicly available, since user documentation is not yet developed, but is described in detail in ref. 46.

**Data availability.** The XSHOOTER observations were made with ESO telescopes at the La Silla Paranal Observatory under Programme 087.B-0812(B), and the data are publicly available at http://archive.eso.org. The analysis also uses HST data from the CLASH survey (programme ID 12460), which is publicly available at http://archive.stsci.edu/hst/. The reduced XSHOOTER two-dimensional spectrum used for the kinematic analysis and the binned one-dimensional spectrum used for the stellar population synthesis analysis is available at https://sid.erda.dk/sharelink/g9s5FnqZjT. The best-fitting stellar population model to the full XSHOOTER spectrum is available on request.

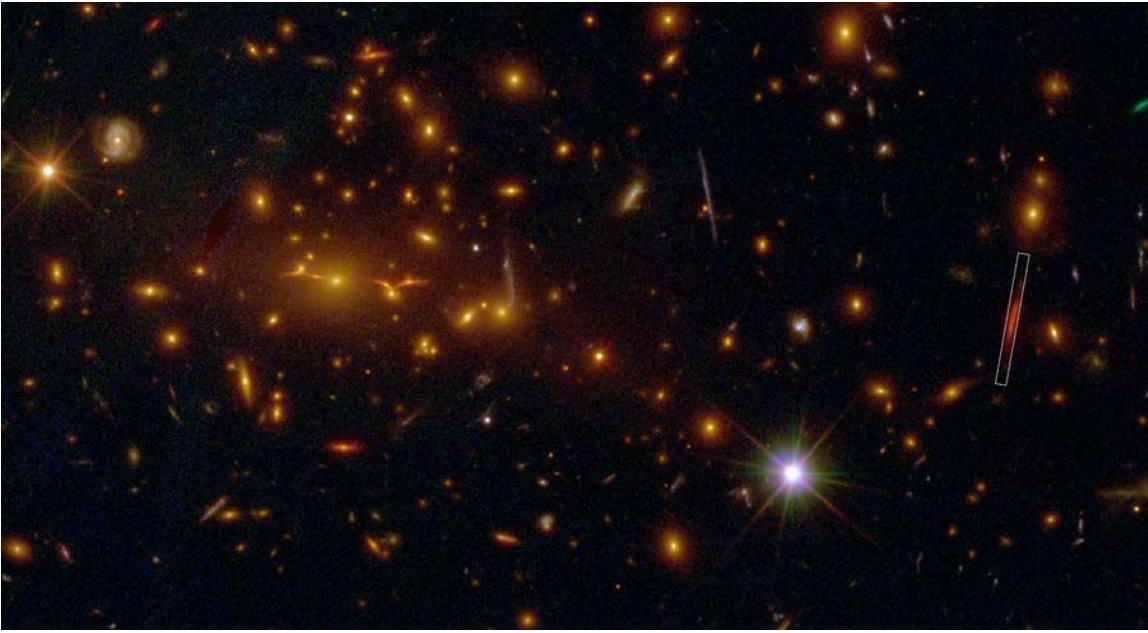

**Extended Data Figure 1** | HST colour-composite image of the lensing cluster MACS2129−1. Indicated is the position of the XSHOOTER slit on the target, which has been magnified and stretched by an average factor of about 4.6 by the foreground cluster. The image is a colour composite (B = F435W + F475W; G = F555W + F606W + F775W + F814W + F850LP; R = F105W + F110W + F125W + F140W + F160W) constructed from CLASH data[63].

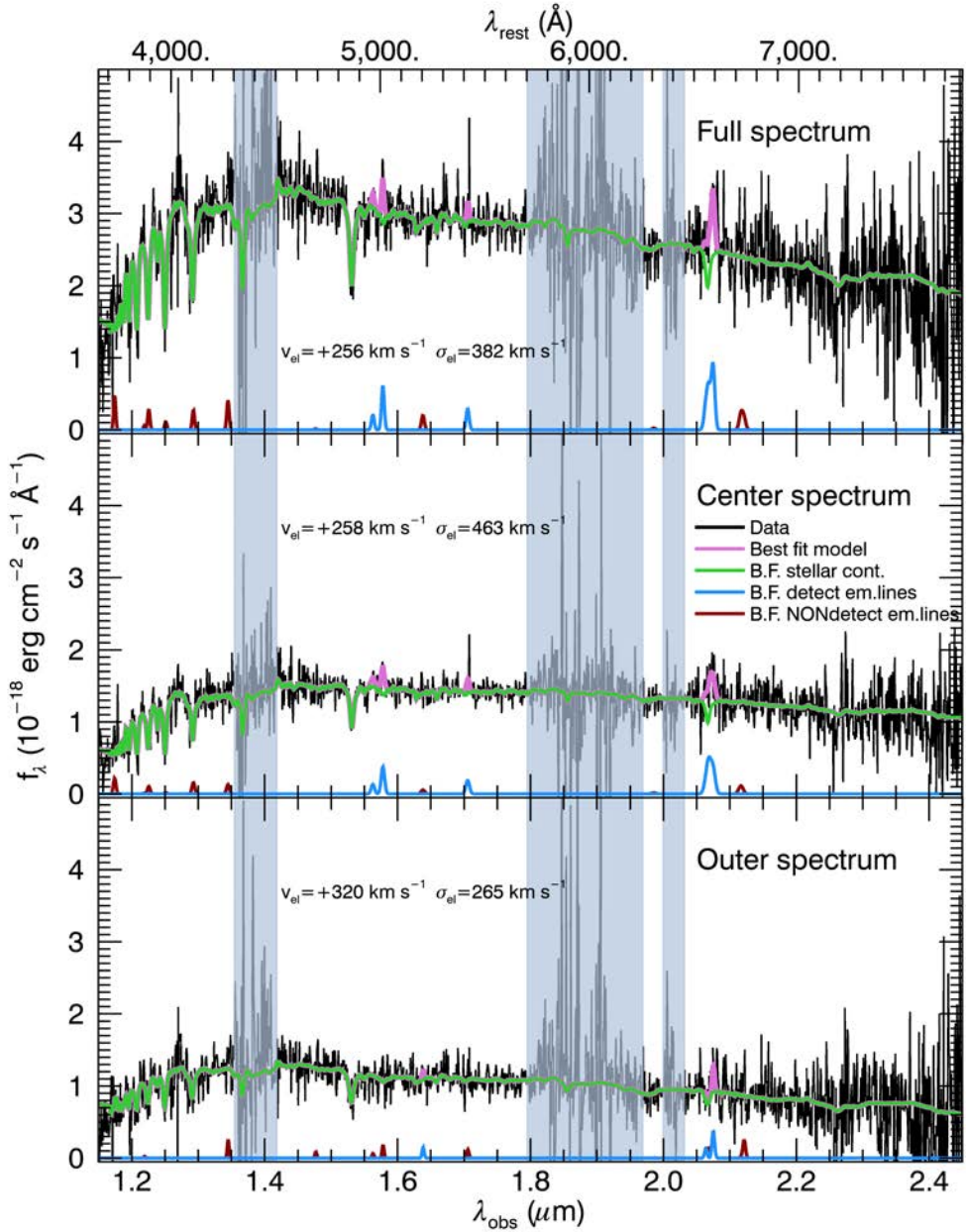

**Extended Data Figure 2 |** Emission line characterization in three spatial extractions of the XSHOOTER spectrum. The top, middle and bottom panels show the full ($|r| < 1.36''$), central ($|r| < 0.5''$) and outer ($0.5'' < |r| < 1.36''$) extractions, respectively, where $|r|$ is the absolute spatial distance to the center of the galaxy. Plotted is the flux-density ($f_\lambda$) versus the observed wavelength ($\lambda_{obs}$) and the rest-frame wavelength ($\lambda_{rest}$). As described in the legend, the coloured lines represent spectral decomposition into nebular emission lines and stellar continuum, obtained with pPXF/GANDALF[50]: the pink line displays the best-fitting composite model; the green line is the best-fitting stellar continuum; the blue and dark red lines represent the best-fitting emission lines with and without a statistically significant detection, respectively. Shaded regions indicate spectral regions of low atmospheric transmission or high background that have been excluded from the fit. On each panel the best-fitting (B.F.) systematic velocity shift $V_{el}$ and dispersion $\sigma_{el}$ of the detected emission (em) lines are indicated.

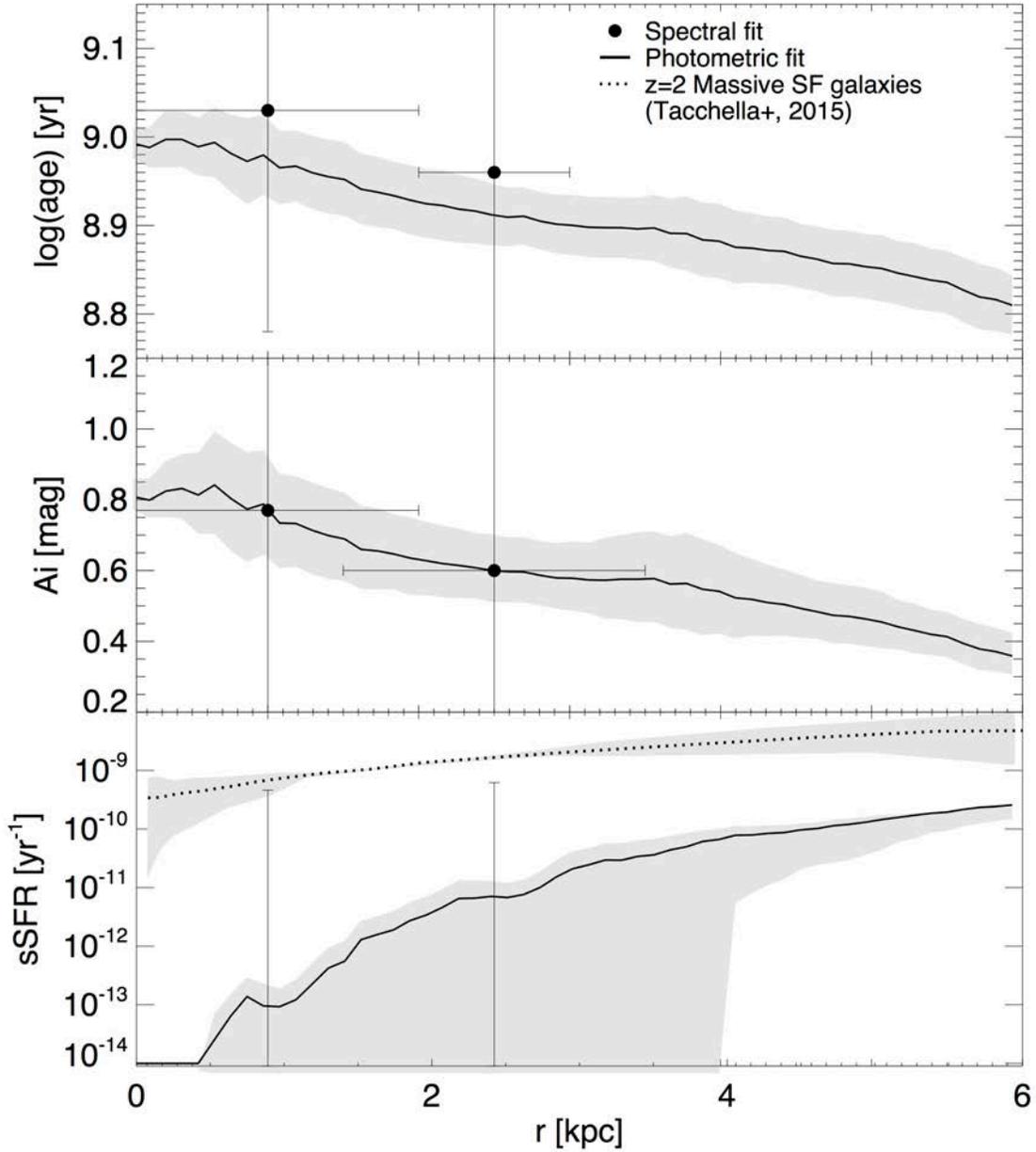

**Extended Data Figure 3 | Radial stellar population gradients.** The full lines show azimuthally averaged radial profiles of median-likelihood stellar population synthesis parameters, derived from the maps in Fig. 3 in elliptical apertures following the best-fitting two-dimensional surface brightness fit. The shaded areas represent the pixel-to-pixel scatter in the median values in the elliptical apertures, not the uncertainties on the individual estimates (see main text). The filled circles with error bars show the median-likelihood parameters and their 68% confidence range from the spectral fits to the central and outer extractions. The dotted line shows the average specific SFR (sSFR) profile from a sample of star-forming (SF) galaxies[22] of mass and redshift similar to that of MACS2129− 1.

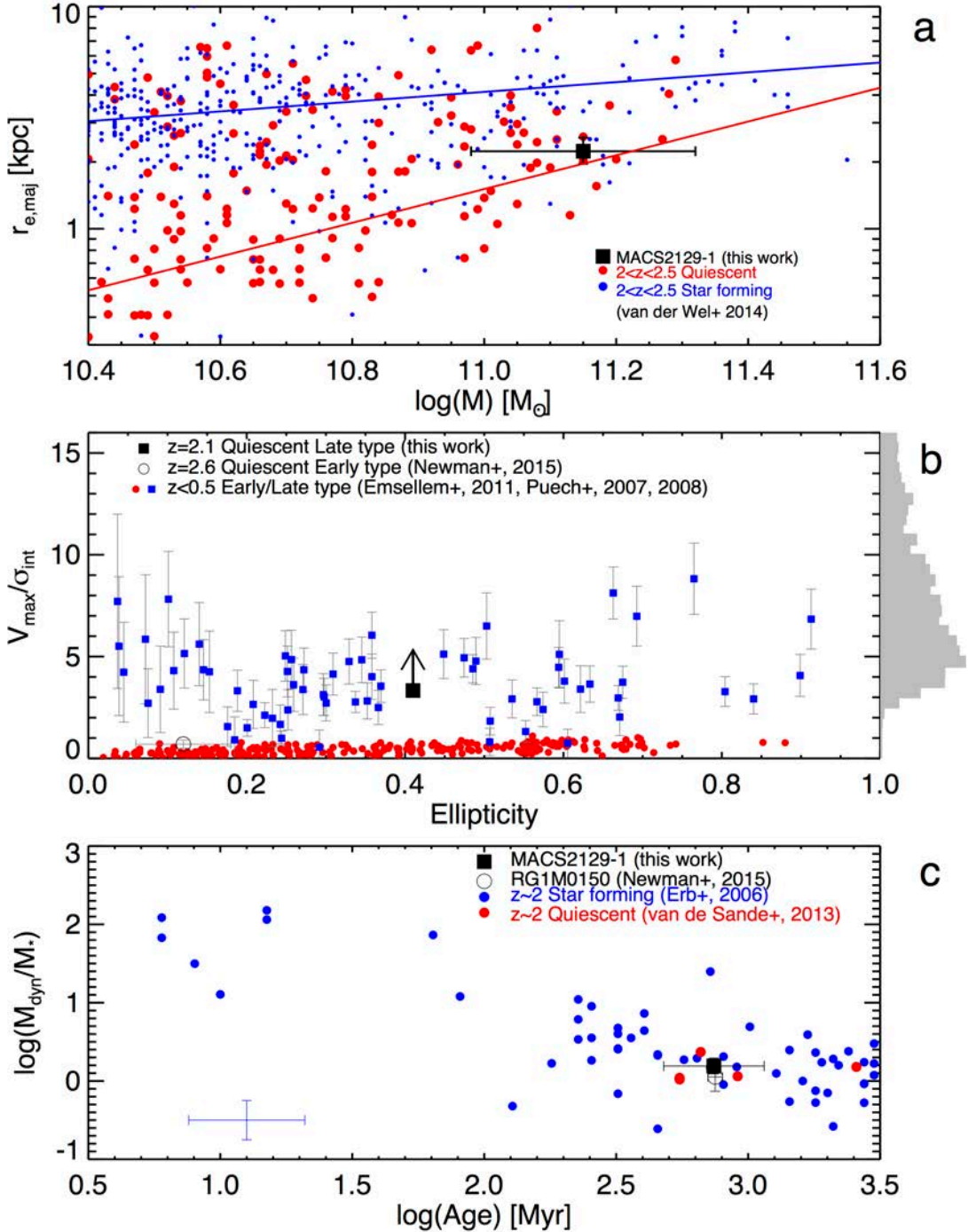

**Extended Data Figure 4 | Properties of MACS2129−1 compared to different galaxy populations.**
**a**, Stellar masses and sizes (major-axis effective radii, $r_{e,maj}$) of $2 < z < 2.5$ galaxies in the CANDELS survey[2]. MACS2129−1 falls on the relation for quiescent galaxies. The error bars include both statistical and systematic errors associated with the fitting added in quadrature. **b**, $V_{max}/\sigma_{int}$ versus ellipticity for the two lensed $z > 2$ compact quiescent galaxies MACS2129− 1 and RG1M0150 (ref. 7) compared to similar-mass local galaxies. The grey histogram shows the V/σ posterior distribution from our modelling. MACS2129− 1 is thus similar to local late types[61,64] (blue), while RG1M0150 is similar to local early types (red). **c,** The dynamical to stellar mass ratio (within re) of MACS2129−1 is similar to previously observed $z > 2$ compact quiescent galaxies, including the strongly lensed RG1M0150, and to $z \approx 2$ starforming galaxies of similar age[49].

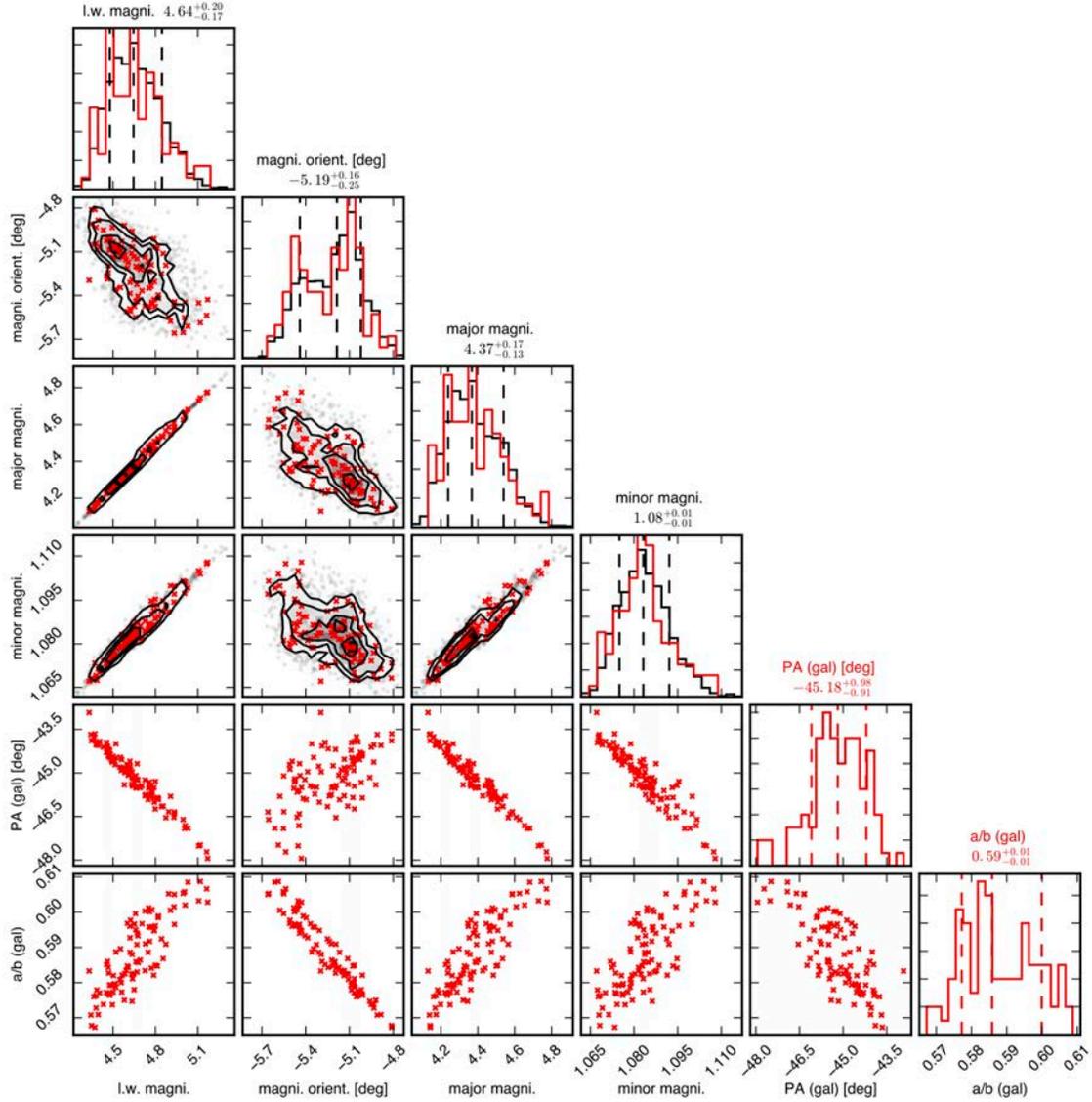

**Extended Data Figure 5 | Correlations between lensing model parameters and derived structural parameters for MACS2129−1.** Shown are the average light-weighted ('l.w.') magnification, the orientation of maximum magnification at the position of MACS2129−1 ('magni. orient.'), the magnification along this axis ('major magni.') and perpendicular to it ('minor magni.'). These were obtained from 1,979 lensing model realizations (black) sampling the full probability distribution. Also shown are correlations with the galaxy ('gal') axis ratios (a/b) and position angles (PA) of MACS2129−1 derived from Galfit analysis of reconstructed source-plane images for a subsample of 98 representative realizations (red).

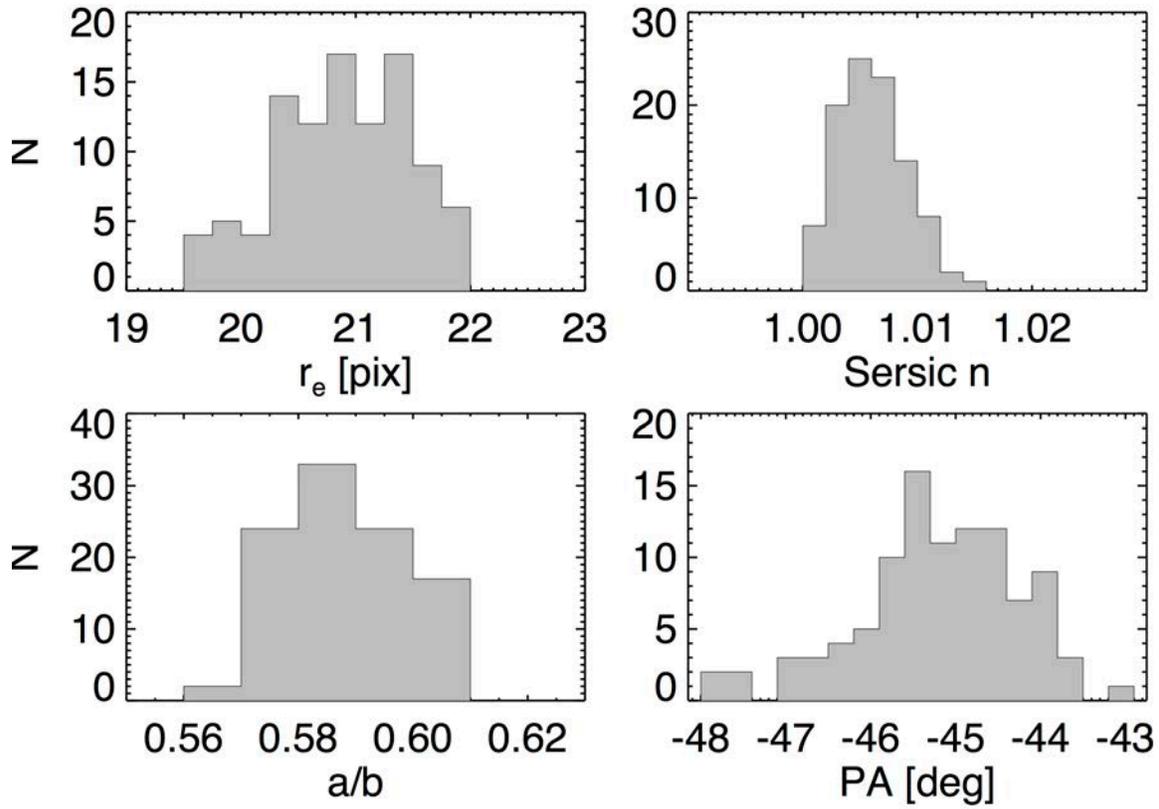

**Extended Data Figure 6 | Structural parameters**. Distributions of the Sersic model parameter n, the effective radius $r_e$, the axis ratio a/b and the position angle PA, derived from two-dimensional surface brightness fits with Galfit, of the source-plane images generated from 98 representative realizations of the lensing model. We adopt the median values of these distributions and their standard deviations as our best-fitting parameters.

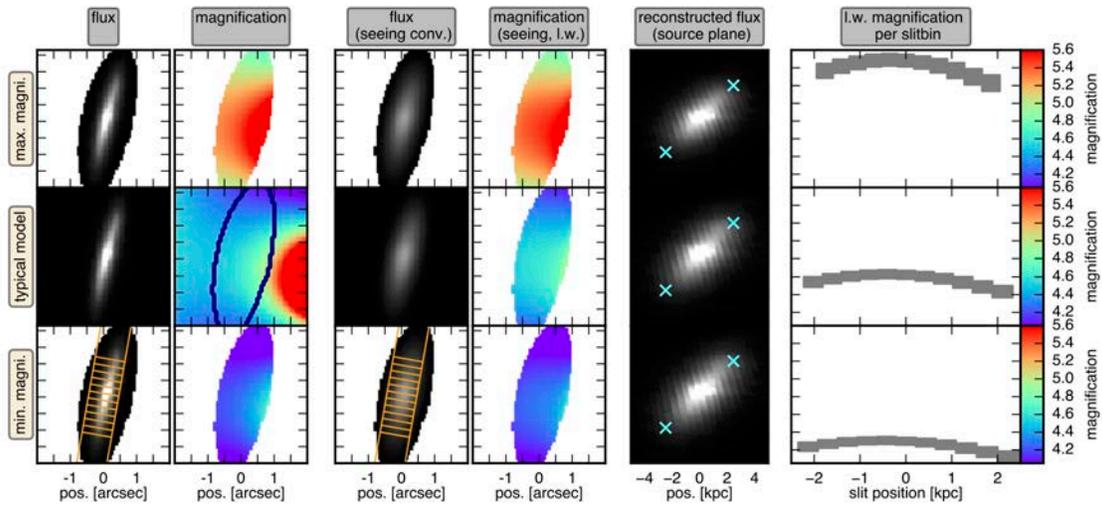

**Extended Data Figure 7 | Variations of the magnification over MACS2129−1.** Results are shown for a typical realization (middle row), and for the realizations with the maximum (top row) and minimum (bottom row) magnifications for different positions (pos.) within the galaxy. The columns (from left to right) show the observed F160W image, the magnification map, the seeing convolved (FWHM = 0.5″) F160W image, the seeing convolved light (F160W)-weighted magnification map, the source-plane image (crosses at same position) and the average light-weighted magnification contributing to each spatial bin in the XSHOOTER slit (shown in the bottom row). The minor variations are caused by the galaxy 3.5″ west of MACS2129−1 (see middle row).

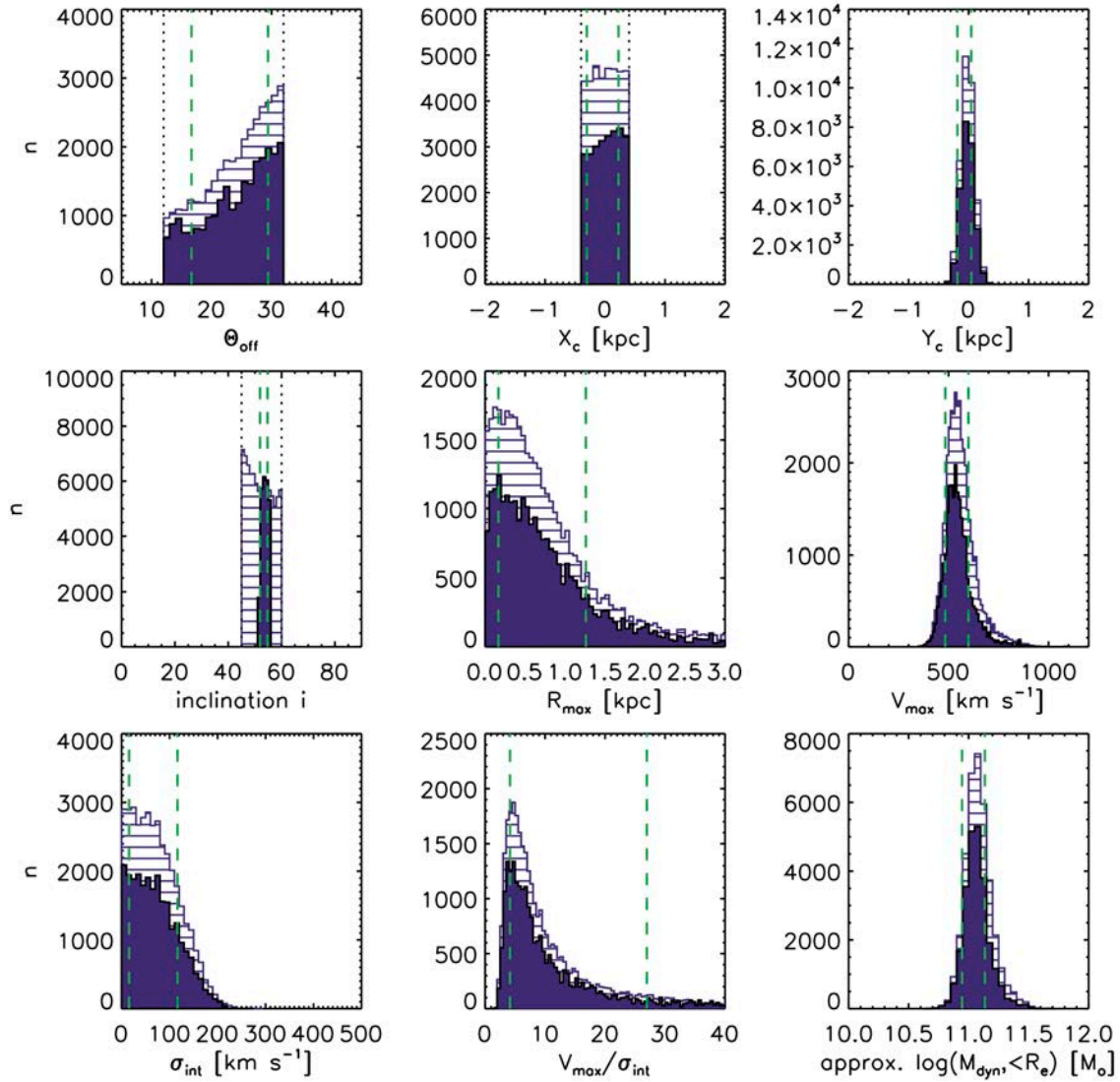

**Extended Data Figure 8 | Posterior distributions for the parameters in our dynamical modelling of the rotation and dispersion curves.** Distributions are shown for the seven free parameters of the model: the offset angle between the slit and the major axis of the disk $\Theta_{off}$, the disk inclination i, the maximum velocity of the disk $V_{max}$, the radius at which the disk reaches $V_{max}$ ($R_{max}$), the position of the centre of the slit relative to the disk centre ($X_c$, $Y_c$), and the intrinsic velocity dispersion $\sigma_{int}$, which is assumed to be constant across the disk. Also shown are inferred distributions for $V_{max}/\sigma_{int}$ and the dynamical mass $M_{dyn}$. The open histograms show the distributions with priors $\Theta_{off} = 22° \pm 10°$ and $|X_c| < 0.4$ kpc. Filled histograms with the additional prior inclination i = 53.8°±2.13°, all derived from Galfit modelling.

**Extended Data Table 1 | Stellar population parameters and emission line fluxes**

**a**

|  | Library A, Fit1 | Library B, Fit1 | Library A, Fit2 | Library B, Fit2 |
|---|---|---|---|---|
| Log(Age/yr) | $8.97^{+0.26}_{-0.25}$ | $9.02^{+0.25}_{-0.25}$ | $8.87^{+0.25}_{-0.19}$ | $8.86^{+0.23}_{-0.17}$ |
| Log($Z/Z_\odot$) | $-0.56^{+0.55}_{-0.55}$ | $-0.58^{+0.54}_{-0.53}$ | $-0.53^{+0.53}_{-0.53}$ | $-0.51^{+0.51}_{-0.56}$ |
| $A_i$ [mag] | $0.60^{+0.90}_{-0.60}$ | $0.33^{+0.88}_{-0.33}$ | $0.73^{+0.73}_{-0.71}$ | $0.70^{+0.53}_{-0.70}$ |
| Log($M_*/M_\odot$) | $11.15^{+0.23}_{-0.20}$ | $11.10^{+0.26}_{-0.20}$ | $11.05^{+0.19}_{-0.16}$ | $11.03^{+0.19}_{-0.16}$ |
| SFR [$M_\odot$/yr] | $0.0^{+139.6}_{-0.0}$ | $0.0^{+0.2}_{-0.0}$ | $0.9^{+86.1}_{-0.9}$ | $0.5^{+14.1}_{-0.5}$ |

**b**

|  | Center (|r|<0.5") | Outer (0.5" < |r| < 1.4") |
|---|---|---|
| Log(Age/yr) | $9.03^{+0.24}_{-0.28}$ | $8.96^{+0.27}_{-0.26}$ |
| Log($Z/Z_\odot$) | $-0.49^{+0.56}_{-0.55}$ | $-0.56^{+0.57}_{-0.56}$ |
| $A_i$ [mag] | $0.77^{+0.94}_{-0.75}$ | $0.60^{+0.98}_{-0.60}$ |
| sSFR [$yr^{-1}$] | $0.0^{+4.6e-10}_{-0.0}$ | $0.0^{+6.2e-10}_{-0.0}$ |

**c**

|  | [OIII]$_{5007}$ [erg/s/cm$^2$] x10$^{-18}$ | HeII$_{5411}$ [erg/s/cm$^2$] x10$^{-18}$ | H$\alpha_{6563}$ [erg/s/cm$^2$] x10$^{-18}$ | [NI]$_{5198}$ [erg/s/cm$^2$] x10$^{-18}$ | [NII]$_{6583}$ [erg/s/cm$^2$] x10$^{-18}$ | $\sigma_{el}$ [km/s] | $V_{el}$ [km/s] |
|---|---|---|---|---|---|---|---|
| Total | 2.8±0.3 | 1.4±0.3 | 3.3±0.8 | 1.0±0.3 | 5.3±0.7 | 382 | 236 |
| Central | 2.1±0.2 | 1.1±0.2 | 2.9±0.5 | - | 2.5±0.5 | 463 | 258 |
| Outer | - | - | - | 0.5±0.2 | 1.5±0.4 | 265 | 320 |

**a**, Median likelihood and 16%–84% confidence intervals for the parameters of our stellar population characterization of MACS2129−1. The age and metallicity (Z) are light-weighted. The SFR is the mean over the last $10^7$ years. The SFR and stellar mass ($M_*$) are corrected for magnification. $A_i$ is the extinction in the rest-frame i-band. sSFR = SFR/$M_*$ is the specific star formation rate. Library A is our default library, while Library B includes only models without additional random bursts of star formation. Fit 1 is performed on the whole XSHOOTER spectrum pixel by pixel, while Fit 2 is performed on absorption-line indices and optical–NIR–MIR photometry. **b**, Median likelihood and 16%–84% confidence intervals for the parameters of our stellar population characterization of MACS2129−1 using Library A/Fit 1, for a central (| r| < 0.5″) and an outer (0.5″ < | r| < 1.4″) extraction, where |r| is the absolute spatial distance to the center of the galaxy. **c**, Detected emission-line fluxes, widths ($\sigma_{el}$) and systematic velocity offsets $V_{el}$ (relative to the absorption lines), in three spatial extractions of the XSHOOTER spectrum (see Extended Data Fig. 2). The total is extracted at | r| < 1.4″.

**Extended Data Table 2 | Dynamical modelling results**

a

| Parameter | Median | 67% Confidence |
|---|---|---|
| $\Theta_{off}$ [deg] | 26.4 | [16.6, 29.4] |
| Inclination [deg] | 53.8 | [51.9, 54.7] |
| $R_{max}$ [kpc] | 0.5 | [0.2, 1.3] |
| $V_{max}$ [km/s] | 532 | [483, 599] |
| $X_c$ [kpc] | 0.0 | [-0.3, 0.2] |
| $Y_c$ [kpc] | 0.0 | [-0.2, 0.0] |
| $\sigma_{intr}$ [km/s] | 59 | [16, 116] |
| $V_{max} / \sigma_{intr}$ | 22 | [4, 27] |
| $Log(M_{dyn}/M_\odot)$ | 11.0 | [10.9, 11.1] |

b

| Model realizations | $V_{max}$ [km/s] | $R_{max}$ [kpc] | $\sigma_{int}$ [km/s] |
|---|---|---|---|
| Benchmark | $532^{+67}_{-49}$ | $0.5^{+0.8}_{-0.3}$ | $59^{+57}_{-44}$ |
| Max magnification | $524^{+52}_{-54}$ | $0.4^{+0.7}_{-0.3}$ | $62^{+57}_{-44}$ |
| Min magnification | $539^{+67}_{-55}$ | $0.5^{+0.8}_{-0.3}$ | $55^{+59}_{-39}$ |
| Max seeing | $543^{+80}_{-50}$ | $0.6^{+0.9}_{-0.4}$ | $49^{+55}_{-36}$ |
| Min seeing | $517^{+51}_{-52}$ | $0.4^{+0.6}_{-0.3}$ | $77^{+51}_{-54}$ |

**a,** Dynamical modelling results from our benchmark model. **b,** Dynamical modelling results, using four different realizations of the seeing and lensing kernels spanning the extreme highest and lowest magnifications found in our 1,979 realizations, and the worst and best seeing allowed by our data. Results are listed for the parameters of the model given above. In all cases the results are derived with the following priors: the offset angle between the slit and the major axis of the disk $\Theta_{off} = 22°\pm10°$, the X position of the centre of the slit relative to the disk centre $|X_c| < 0.4$ kpc and disk inclination i = 53.8°±2.13°.